\documentclass[useAMS,usenatbib]{mn2e}
\usepackage{graphicx}
\title[The solar flare-induced sunquake of 2005 January 15]{Helioseismic analysis of the solar flare-induced sunquake of 2005 January 15}
\author[H. Moradi, A. -C. Donea, C. Lindsey, D. Besliu-Ionescu and P. S. Cally]{H. Moradi$^{1}$\thanks{E-mail:
hamed.moradi@sci.monash.edu.au}, A. -C. Donea$^1$, C. Lindsey$^2$, D. Besliu-Ionescu$^{1,3}$ and P. S. Cally$^1$\\
$^{1}$Centre for Stellar and Planetary Astrophysics, School of 
Mathematical Sciences, Monash University, Victoria 3800, Australia\\
$^{2}$Colorado Research Associates Division, NorthWest Research 
Associates Inc., 3380 Mitchell Lane, Boulder, CO, 80301, U.S.A\\
$^{3}$Astronomical Institute of the Romanian Academy, RO-040557, 
Bucharest, Romania\\}

\begin{document}


\maketitle

\begin{abstract}
We report the discovery of one of the most powerful sunquakes
detected to date, produced by an X1.2-class solar flare in active
region 10720 on 2005 January 15.  We used helioseismic holography to
image the source of seismic waves emitted into the solar interior from
the site of the flare.  Acoustic egression power maps at 3 and 6 mHz
with a 2~mHz bandpass reveal a compact acoustic source strongly
correlated with impulsive hard X-ray and visible-continuum emission
along the penumbral neutral line separating the two major opposing
umbrae in the $\delta$-configuration sunspot that predominates
AR10720.  At 6~mHz the seismic source has two components, an intense,
compact kernel located on the penumbral neutral line of the
$\delta$-configuration sunspot that predominates AR10720, and a
significantly more diffuse signature distributed along the neutral
line up to $\sim$15~Mm east and $\sim$30~Mm west of the kernel.  The
acoustic emission signatures were directly aligned with both hard
X-ray and visible continuum emission that emanated during the flare.
The visible continuum emission is estimated at $2.0 \times 10^{23}$~J,
approximately 500 times the seismic emission of $\sim 4 \times
10^{20}$~J. The flare of 2005 January 15 exhibits the same close
spatial alignment between the sources of the seismic emission and
impulsive visible continuum emission as previous flares, reinforcing
the hypothesis that the acoustic emission may be driven by heating of
the low photosphere. However, it is a major exception in that there
was no signature to indicate the inclusion of protons in the particle
beams thought to supply the energy radiated by the flare.  The
continued strong coincidence between the sources of seismic emission
and impulsive visible continuum emission in the case of a
proton-deficient white-light flare lends substantial support to the
``back\,--\,warming'' hypothesis, that the low photosphere is
significantly heated by intense Balmer and Paschen continuum-edge
radiation from the overlying chromosphere in white-light flares.

\end{abstract}


\section{Introduction}

Although most large solar flares appear to be acoustically inactive,
certain energetic flares radiate intense seismic transients into the
solar interior during the impulsive phase.  These wave packets radiate
thousands of kilometres from the flaring region into the solar
interior, but most of this energy is refracted back to the solar
surface within approximately 50~Mm of the source and within an hour of
the beginning of the flare.  The surface manifestation is a
wave-packet of ripples accelerating outward from the general source
region that is sometimes obvious in raw helioseismic observations.
\citet{kz1998} discovered the first known instance of seismic
emission, from the X2-class flare of 1996 July 9 in AR7978,
identifying the phenomenon by the name ``sunquake.''

For a long time these events were thought to be an extremely rare phenomenon. 
However, with the advancement of local helioseismic techniques such as 
helioseismic holography \citep{lb2000}, we have now detected numerous seismic 
sources of varying intensity produced by X- and high M-class flares 
\citep{dbl1999,dl2005,donea2005}. 

A subsequent extensive survey of X-class solar flares (Besliu-Ionescu et al., in preparation)
led to the discovery of more than a dozen seismic emission signatures
from flares.  Almost all of these have occurred in complex active
regions.  In this paper we report on the discovery of one of the most
powerful flare seismic transients detected to date and compare the
acoustic signatures of this sunquake with other supporting
observations.

AR10720 was a complex active region that appeared on the solar disk on
2005 January 11 and soon became one of the largest and most active
sunspot regions of the current solar cycle.  In the period January 15 -- 20, 
AR10720 produced 5 X-class solar flares, including an X7.1 on
January 20, which produced an intense solar proton storm.  However,
helioseismic observations sufficient to show seismic emission were
acquired only for the X1.2 flare of January 15.  This flare was
observed by numerous space and ground-based solar observatories,
including the Michelson Doppler Imager (MDI) instrument on board the Solar and Heliospheric Observatory (\textit{SOHO}), the Reuven
Ramaty High-Energy Solar Spectroscopic Imager (\textit{RHESSI}), the 
Geostationary Operational Environmental Satellite (\textit{GOES}),
the Transition Region and Coronal Explorer (\textit{TRACE}), and the
earth-based Global Oscillations Network Group (GONG).
AR10720 itself was observed by the Imaging Vector Magnetograph (IVM)
at the Mees Solar Observatory in the general time frame of the 15
January 2005 flare.

In Section 2 we present a brief description of the helioseismic observations
we analyzed and in Section 3 we review the technique of computational seismic 
holography. 
Section 4 consists of our results and analysis, and finally in Section 5 we 
present a discussion and summary.      

\section[]{The Helioseismic Observations}

The MDI data consist of full-disk Doppler images in the photospheric line 
 Ni~I~6768~\AA, obtained at a cadence of 1 minute, in addition to approximately
hourly continuum intensity images and line-of-sight magnetograms. 
The MDI data sets are described in more detail by \citet{setal1995}. 
For the flare of 2005 January 15, we analyzed a dataset with a period of 
four hours encompassing the flare. 
For the purpose of our analysis, the MDI images we obtained
(Dopplergrams, magnetograms and intensity continuum) were remapped onto 
a Postel projection \citep{deforest2004} that tracks solar rotation, 
with the region of interest fixed at the centre of the projection. 
The nominal pixel separation of the projection was 0.002 solar radii 
(1.4~Mm) with a $256\times256$ pixels field of view, thus encompassing 
a region of approximately $360 \times 360$~Mm$^2$ on the solar surface. 

\section{Helioseismic Holography}

We briefly review the adaptation of computational seismic holography
for applications in flare seismology.  In general, helioseismic
holography is the phase-coherent reconstruction of acoustic waves
observed at the solar surface into the solar interior to render
stigmatic images of subsurface sources that have given rise to the
surface disturbance.  Because the solar interior refracts down-going
waves back to the surface, helioseismic holography can likewise use
observations in one surface region, the pupil, to image another, the
focus, a considerable distance from the pupil.  We call this ``seismic
holography from the subjacent vantage'' (see fig. 4 of
\citet{lb2000}).  The subjacent vantage renders the photosphere
as viewed by an acoustic observer directly beneath it.

In general the acoustic reconstruction can be done either forward or
backward in time.  When it is backward in time, we call the
extrapolated field the ``acoustic egression.''  In the case of
subjacent vantage holography, this represents waves emanating from the
surface focus downward into the solar interior.  When the surface
acoustic field at any point $\textbf{r}'$ in the pupil is expressed as
a complex amplitude $\hat{\psi}$ for any given frequency $\omega$, the
acoustic egression can be expressed as
\begin{equation}
\hat H_+(\textbf{r}, ~\omega)
 ~=~ \int_{pupil} \hat{G}_+(\textbf{r}, ~\textbf{r}', ~\omega)
 ~\hat{\psi}(\textbf{r}', ~\omega)d^2\textbf{r}'.
\label{eq:Hpluscalc}
\end{equation}

In this formalism, 
$\hat{G}_+(\textbf{r}, ~\textbf{r}',~\omega)$ is a Green's function that 
expresses the disturbance at the focus, $\textbf{r}$, due to a measured point 
source at surface point $\textbf{r}'$ from which the acoustic wave is supposed
to propagate backwards in time to the focus. 

The relation between the complex amplitude, $\hat\psi(\textbf{r},~\omega)$, 
of frequency appearing in equation (\ref{eq:Hpluscalc}) and the real acoustic 
field, $\psi(\textbf{r},~t)$, representing the surface acoustic field in the 
MDI observations as a function of time is expressed by the Fourier transform:
\begin{equation}
\psi(\textbf{r}, ~t)
 ~=~ \frac{1}{\sqrt{2\pi}}\int^{\infty}_{-\infty} e^{i\omega t}
 ~\hat{\psi}(\textbf{r}, ~\omega) ~d\omega.
\label{eq:psit}
\end{equation}
The same applies to the acoustic egression:
\begin{equation}
H_+(\textbf{r}, ~t)
 ~=~ \frac{1}{\sqrt{2\pi}}\int^{\infty}_{-\infty}e^{iwt}
 ~\hat{H}_+(\textbf{r}, ~\omega) ~d\omega.
\label{eq:Hplust}
\end{equation}
The ``egression power",
\begin{equation}
P(\textbf{r}, ~t) ~=~ |H_+(\textbf{r}, ~t)|^2,
\label{eq:Hpwr}
\end{equation}
is used extensively in holography of acoustic sources and absorbers.
Equation (\ref{eq:Hpwr}) is used to produce egression power maps, which show 
compact positive signatures in the spatial and temporal neighbourhoods of 
localized seismic transient emitters.
The signature of a localized absorber illuminated by ambient acoustic noise 
is a similarly sharp deficit in egression power, appearing as a silhouette 
against a generally positive background when rendered graphically. 

In this study $P({\bf r}, t)$ is separately derived from computations
of $\hat H_+({\bf r}, \omega )$ over 2\,--\,4~mHz and 5\,--\,7~mHz
ranges of the spectrum.  In practice, there are major diagnostic
advantages to the 5\,--\,7~mHz spectrum, as it avoids the much greater
quiet-Sun ambient noise at lower frequencies, which competes
unfavourably with acoustic emission into the pupil from the flare.  Due
to a shorter wavelength, the high frequency band also provides us with
waves that have a finer diffraction limit.  These advantages come at
some expense in temporal discrimination, as the egression power
signatures that result are temporally smeared to a minimum effective
duration of order
\begin{equation}
\Delta t ~=~ \frac{1}{\Delta\nu} ~=~ \frac{1}{2~{\rm mHz}} ~=~ 500~{\rm s}.
\label{eq:Deltat}
\end{equation}
This smearing operates in both directions in time, meaning that the acoustic 
signature of the flare $P({\bf r}, t)$, once the computation is complete, will 
invariably commence several minutes before the actual onset of the flare and 
last for several minutes afterward even if the actual acoustic disturbance was 
instantaneous (and no seismic signature at all reached the pupil until nearly 
twenty minutes after the flare began). 
\begin{figure*}
\begin{center}
\includegraphics[width=0.7\textwidth]{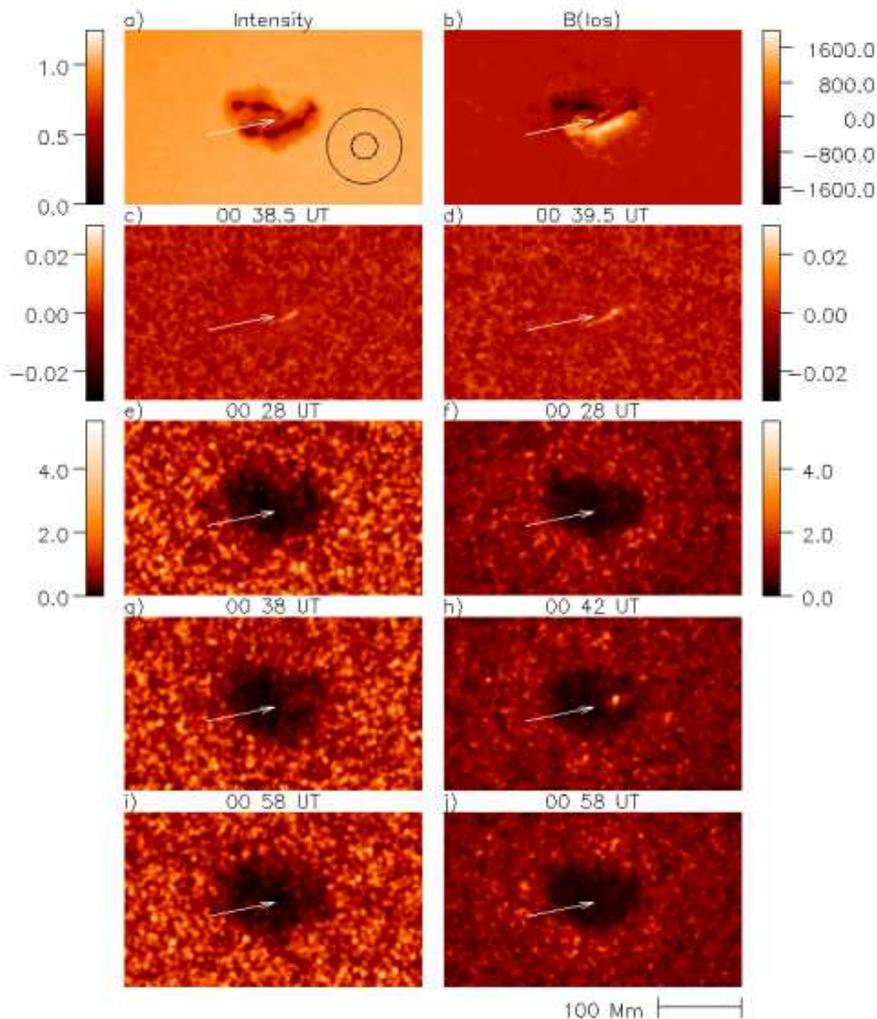}
\caption{Egression power snapshots of AR10720 on 2005 January 15 taken
before, during and after the flare and integrated over a 5\,--\,7~mHz
frequency band.  The top frames show an MDI visible continuum image of
AR10720 (left) at 00:00~UT and a magnetogram (right) at 00:28~UT.
The second row shows continuum intensity differences 30 seconds before and
after the time that appears above the respective frames, taken from
the GONG observatory at Mauna Loa.  The bottom three rows show egression
power maps before (row 3), during (row 4), and after (bottom row) the
flare at 3~mHz (left column) and 6~mHz (right column).  The annular
pupil for the egression computations is drawn in the top left panel.
To improve statistics, the original egression power snapshots are
smeared by convolution with a Gaussian with a $1/e$-half-width of
3~Mm.  Times are indicated above respective panels, with arrows
inserted to indicate the location of the acoustic source.  Colour
scales at right and left of row 3 apply to respective columns in rows
3\,--\,5.  Egression power images and the continuum images are
normalized to unity at respective mean quiet-Sun values.  At 3~mHz
this is $\sim$2~kW~m$^{-2}$.  At 6~mHz it is 70~W~m$^{-2}$.}
\label{fig:egpwrsnaps}
\end{center}
\end{figure*}

\section{Results and Analysis}

\subsection{The Seismic Signatures} 

AR10720 was predominated by a single 
$\delta$-configuration sunspot.  The top row of
Fig.~\ref{fig:egpwrsnaps} shows continuum intensity (left) and
line-of-sight magnetic field (right) of the active region shortly
before the flare.  The 2005 January 15 solar flare in AR10720 was
classified as X1.2, localized at N14E08 on the solar surface.  The
\textit{GOES} satellite measured a $1.2\times10^{-1}$Jm$^{-2}$ X-ray flux in
the 1-8~\AA~range integrated over the duration of the flare.  Excess
X-ray emission began at 00:22~UT, reaching a maximum at 00:43~UT, and
ending at 01:02~UT. There was significant white-light emission
with a sudden onset, as indicated by the intensity difference
signatures shown in the second row of Fig.~\ref{fig:egpwrsnaps}, and
this coincided closely with hard X-ray (HXR) signatures indicating high-energy
particles accelerated into the chromosphere.  However, unlike the
flares of 2003 October 28\,--\,29 \citep{dl2005}, there were no
signatures to indicate the inclusion of high-energy protons in these
particle influxes.
\begin{figure*}
\begin{center}
\includegraphics[width=0.7\textwidth, angle=0]{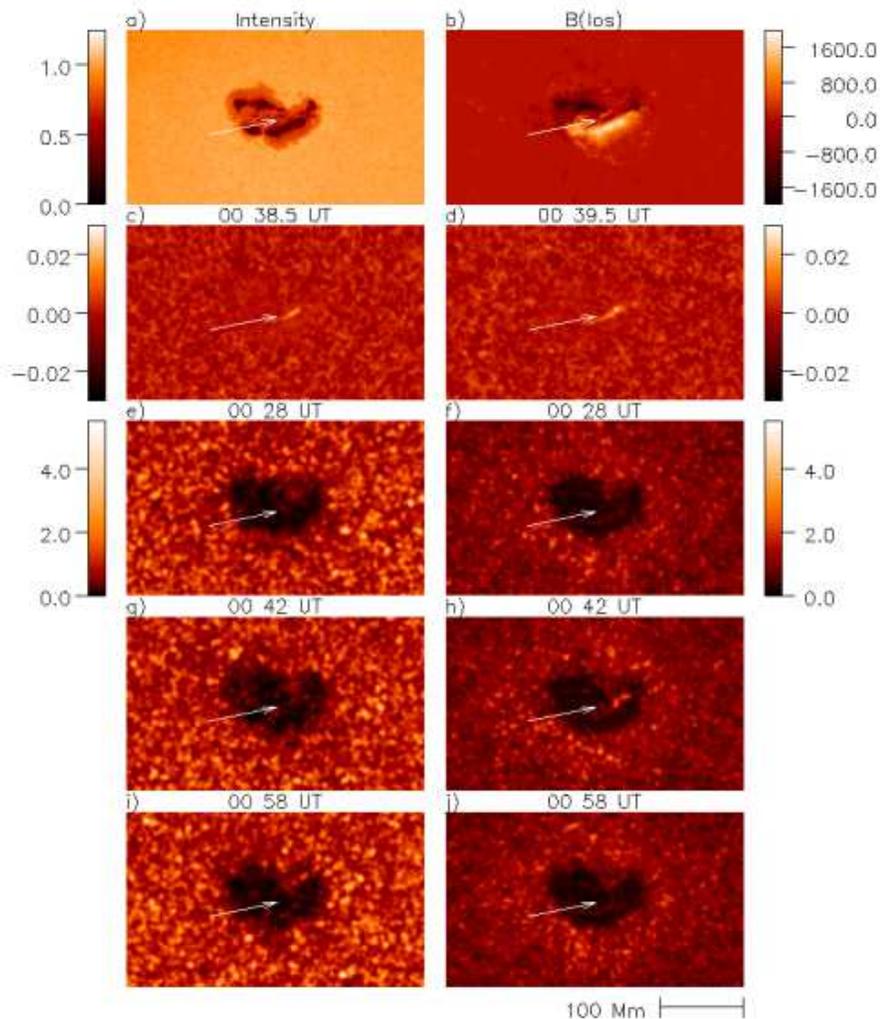}
\caption{Acoustic power snapshots of AR10720 on 2005 January 15.
Details are the same as for Fig.~\ref{fig:egpwrsnaps}, but local acoustic power maps
appear in the bottom three rows in place of egression power maps.}
\label{fig:acpwrsnaps}
\end{center}
\end{figure*}

To assess seismic emission from the flare, we computed the egression,
$H_+$, as prescribed by equation~(\ref{eq:Hplust}) over the
neighbourhood of the active region at one-minute intervals in $t$,
mapping the egression power, $P$, as prescribed by
equation~(\ref{eq:Hpwr}), for each value of $t$.  We call a map of $P$
evaluated at any single $t$ an egression power ``snapshot.''  From
this point will refer to the 5\,--\,7~mHz bandpass simply as 6~mHz and
to the 2\,--\,4~mHz bandpass as 3~mHz.  Egression power snapshots
before, during and after the flare are shown in the bottom three rows
of Fig.~\ref{fig:egpwrsnaps} at 3~mHz (left column) and 6~mHz (right
column). In these computations the pupil was an annulus of radial
range 15\,--\,45~Mm centred on the focus
(Fig.~\ref{fig:egpwrsnaps}a).

All egression power snapshots mapped in Fig.~{\ref{fig:egpwrsnaps}} show 
considerably suppressed acoustic emission from the magnetic region, attributed to 
strong acoustic absorption by magnetic regions, discovered by \citet{bdl1988} (see also \citet{braun1995,blff1998,bl1999a}). Furthermore, all 6~mHz egression 
power snapshots in Fig.~{\ref{fig:acpwrsnaps}} also show acoustic emission ``halos,'' i.e. significantly 
enhanced acoustic emission from the outskirts of complex active regions
\citep{bl1999b,dbl1999}. 

A conspicuous seismic source is seen in the 6~mHz egression power snapshot
at 00:42~UT, whose location is indicated by an arrow in all of the frames.
A close examination of the source shows that it has two components.
By far the most conspicuous component is an intense, compact kernel 
$\sim$10~Mm located on the penumbral neutral line of the $\delta$-configuration
sunspot.
Somewhat more diffuse but clearly significant is a secondary, somewhat
lenticular signature distributed along the neutral line out to $\sim$15~Mm 
east and $\sim$30~Mm west of the kernel.
These signatures correspond closely with other compact manifestations of
the flare.
The kernel accounts for approximately 45 per cent of the egression power integrated 
over the region encompassing the flare signature, with the lenticular 
component outside of the kernel accounting for the rest. 

\begin{figure*}
\begin{center}
\includegraphics[width=0.65\textwidth, angle=0]{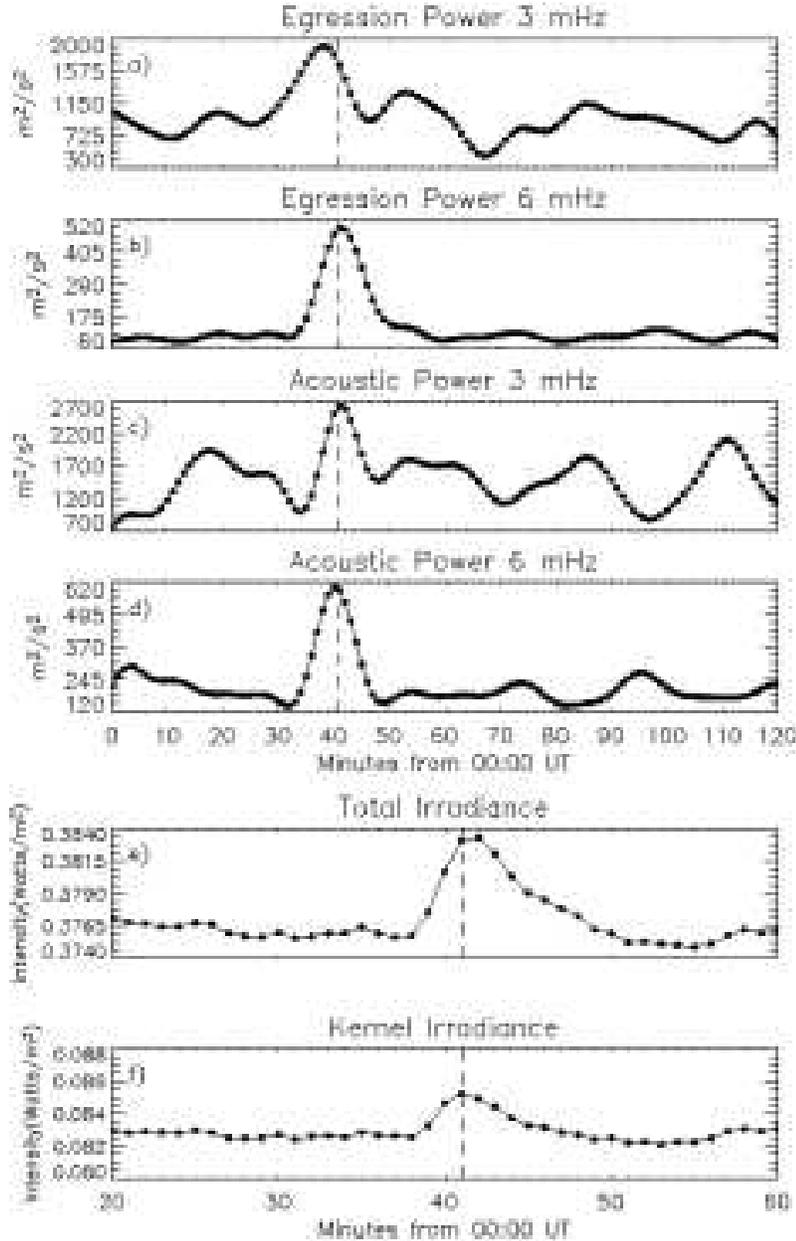}
\caption{Time series of the 3 and 6 mHz egression and acoustic power (integrated 
over the neighbourhood of the egression power signatures) are plotted in the top 
four rows. The dashed vertical lines mark the time of maximum acoustic emission 
(00:41~UT) at 6~mHz. The relatively extended duration of the acoustic signatures 
is a result of limits to temporal resolution imposed by truncation of the spectrum 
(see equation \ref{eq:Deltat}). The bottom two rows show visible continuum irradiance 
at 1~au from the flaring region along the neutral line in the neighbourhood of the 
flare. The emission from the neighbourhood of the kernel component of the 6~mHz acoustic 
source (plot f) is discriminated from the total (plot e).}
\label{fig:plots}
\end{center}
\end{figure*}
 
The 3~mHz egression power snapshots shown in the left column of 
Fig.~\ref{fig:egpwrsnaps} actually show a considerably stronger seismic emission 
signature than the 6~mHz signature (right column). But, because of the much greater 
ambient acoustic emission at this frequency, the 3~mHz signature is not nearly as 
conspicuous or significant as the 6~mHz signature.
It appears to have only a diffuse lenticular component and no conspicuous kernel to 
match the 6~mHz kernel.

It is important to distinguish between the egression power,
$|H_+(\textbf{r}, ~t)|^2$, and the local acoustic power,
$P(\textbf{r}, ~t)$, which is the square modulus, $|\psi(\textbf{r},
~t)|^2$, of the local wave amplitude $\psi$ at the focus,
$\textbf{r}$.  Each pixel in a local acoustic power map represents
local surface motion as viewed directly from above the photosphere.
Each pixel in the egression power map computed by subjacent vantage
holography of the surface is a coherent representation of acoustic
waves that have emanated downward from the focus, deep beneath the
solar surface, and re-emerged into a pupil (see diagram of annulus in
Fig.~\ref{fig:egpwrsnaps}a) 15\,--\,45~Mm from the focus.
\begin{table*}
\centering
\begin{minipage}{120mm}
\caption{Energy estimates of the seismic signatures of sunquakes detected 
to date}
\vspace{1em}
\renewcommand{\arraystretch}{1.2}
\begin{tabular}[thb]{lrcccc}
\hline
Date & Class & 3~mHz &  6~mHz  & 1\,--\,8~\AA ~X-Rays & Visible \\
     & & (ergs) & (ergs) & (ergs) & (ergs) \\
\hline
1996 Jul 09 
& X2.6
& $7.5 \times 10^{27}$
& $2.4 \times 10^{26}$
& $2.8 \times 10^{29}$
& -------------- \\
2001 Sep 09
& M9.5
& $1.1 \times 10^{27}$
& $2.0 \times 10^{26}$
& $6.2 \times 10^{28}$
& $1.2 \times 10^{30}$ \\
2003 Oct 28 
& X17.2
& $4.7 \times 10^{27}$
& $9.4 \times 10^{26}$
& $5.0 \times 10^{30}$
& -------------- \\
2003 Oct 29 
& X10.0
& $9.4 \times 10^{26}$
& $3.5 \times 10^{26}$
& $1.5 \times 10^{30}$
& $3.8 \times 10^{29}$ \\
\textbf{2005 Jan 15}
& \textbf{X1.2}
& \textbf{$2.4 \times 10^{27}$}
& \textbf{$1.0 \times 10^{27}$}
& \textbf{$3.4 \times 10^{29}$}
& \textbf{$2.0 \times 10^{30}$} \\
\hline \\
\end{tabular}
\end{minipage}
\label{tab:table2}
\end{table*}

Fig.~\ref{fig:acpwrsnaps} shows local acoustic power snapshots of AR10720 at 
3~mHz (left column) and 6~mHz (right column) before, during, and after the flare. 
As in the case of egression power (Fig.~\ref{fig:egpwrsnaps}) all of the local 
acoustic power maps show a broad acoustic deficit marking the magnetic region. 
An enhanced local acoustic power halo surrounding the active region is clearly 
apparent in the 6~mHz snapshots. 
The acoustic signature of the flare is also clearly visible at 6~mHz.
This appears to consist of a pair of kernels, a relatively stronger one nearly 
coinciding in location with, but slightly east of, the 6~mHz egression power 
kernel and a weaker one $\sim$10~Mm to west and slight north, lying on the neutral 
line along which the lenticular component of the 6~mHz egression power is 
distributed.
As in the corresponding egression power snapshot, the 3~mHz local acoustic 
power snapshots show a stronger but still less conspicuous signature than 
that at 6~mHz due to a similarly much greater background acoustic power at 
3~Mm.
\begin{figure*}
\begin{center}
\includegraphics[width=0.6\textwidth]{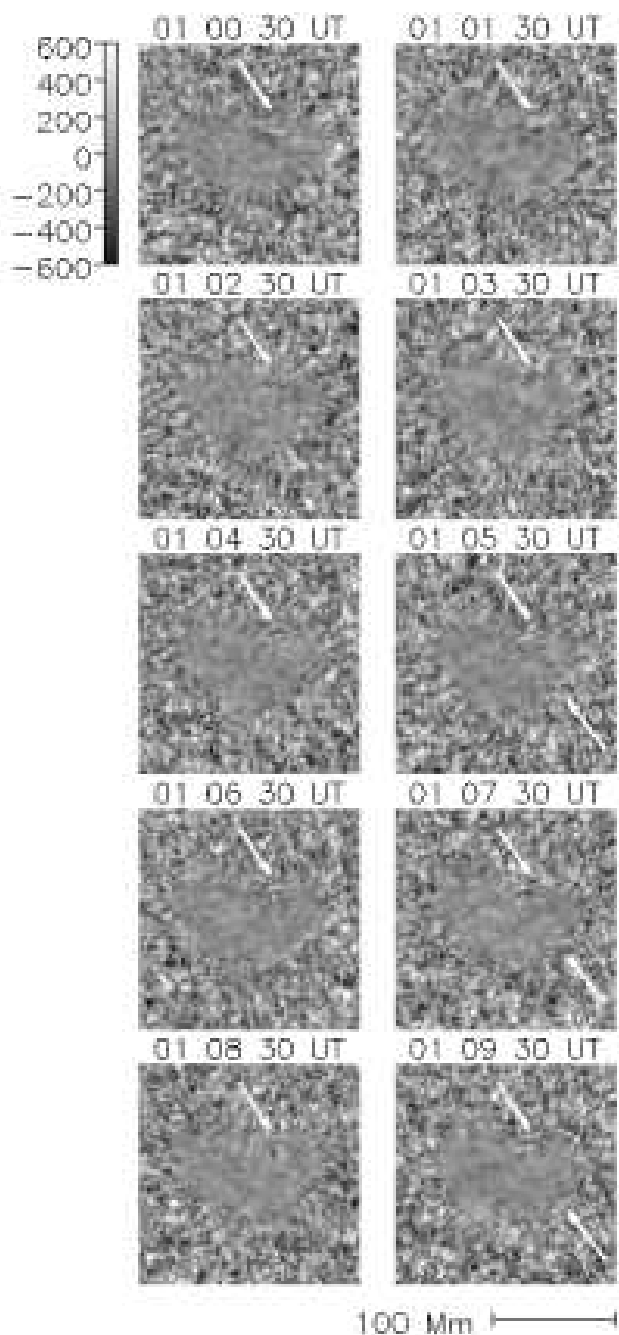}
\caption{MDI Doppler-difference images showing the expanding ring-shaped wave 
packet produced by the 2005 January 15 flare. 
The times shown above respective panels represent one-minute differences. 
The arrows pointing in the South--East direction (i.e. upper arrows) show 
the location of the upper-half of the wave front while arrows pointing in 
the North--West direction (i.e. lower arrows) indicate the lower half of 
the wave front.
Grey-scale at top left expresses Doppler velocity differences in units of
ms$^{-1}$ and applies to all frames in the figure.}
\label{fig:ripples1}
\end{center}
\end{figure*}

Fig.~\ref{fig:plots} shows plots of the egression and acoustic power time 
series in the 3 and 6 mHz bands and continuum emission in the neighbourhood
of the seismic signature, discriminating continuum emission in the region of the
kernel component in the 6~mHz egression power signature from the total.
The flare irradiance profiles were extrapolated by applying the assumption that 
the irradiance is directly proportional to the GONG continuum signature in
the neighbourhood of Ni~I~6768~\AA \citep{dl2005}.
\begin{figure*}
\begin{center}
\includegraphics[width=0.9\textwidth]{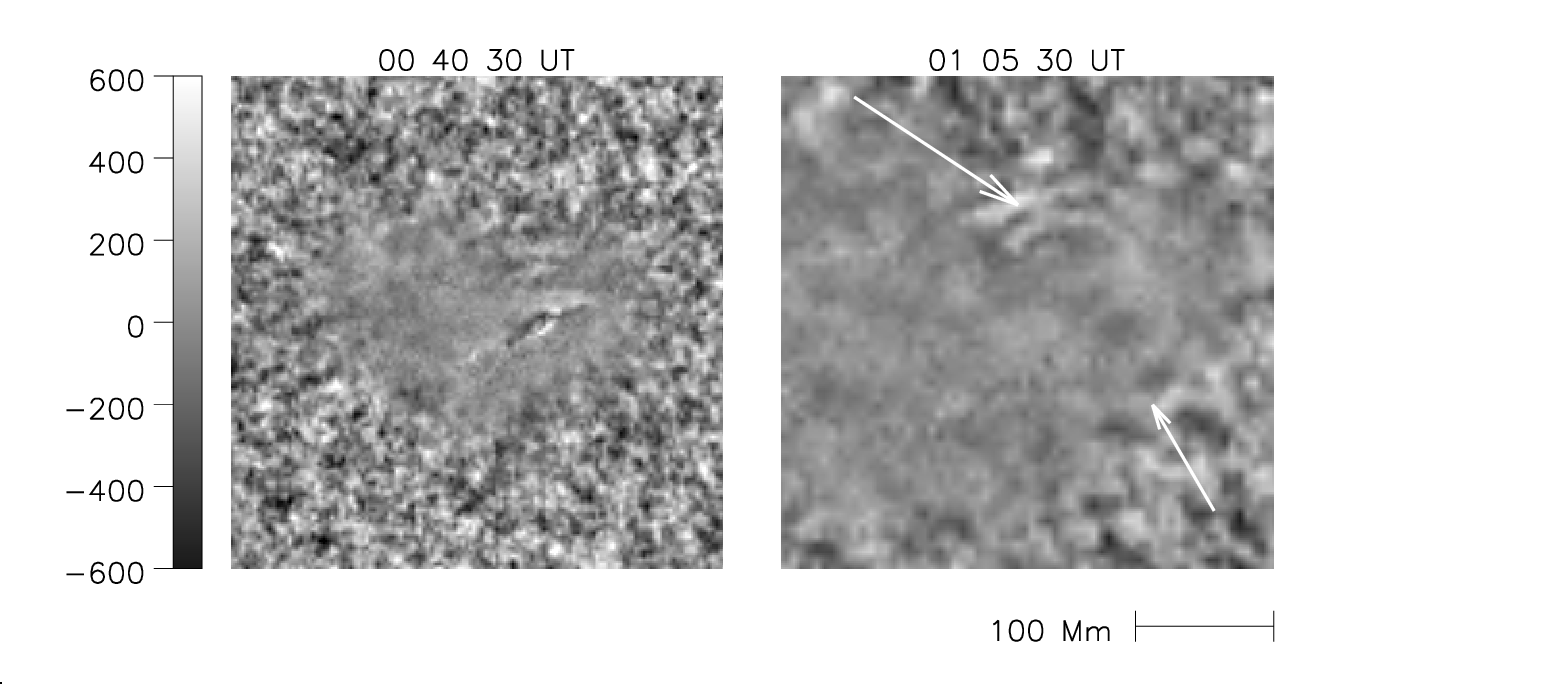}
\caption{MDI Doppler-difference images showing ring-shaped wave packet 
accelerating outward from the site of the 2005 January 15 flare.
The left panel shows the local Doppler signature along the magnetic
neutral line during the impulsive phase of the flare at 00:40:30~UT.
Arrows in the right panel indicate locations of the ripples 
propagating outwards from the site of local disturbance 25 minutes later.}%
\label{fig:ripples2}
\end{center}
\end{figure*}

The flare of 2005 January 15 produced the most conspicuous acoustic signature 
of any flare that has produced a detectable seismic emission. 
This appears to be because such a large fraction of the energy was released 
into the high-frequency (5\,--\,7~mHz) spectrum, where the competing ambient
acoustic power is so far suppressed.
Table \ref{tab:table2} shows the energy estimates of the seismic transients radiated
into the active region subphotosphere by five flares that have produced 
conspicuous seismic signatures%
\footnote{The energy estimates in Table \ref{tab:table2} were obtained by integrating 
the egression power over the neighbourhood of the seismic sources (e.g. those shown 
in Figs.~1g,h for the 2005 January 15 flare).
This computation is blind to waves that miss the 15\,--\,45~Mm in the first skip.
Comparative seismic holography applied to simulated acoustic transients, and to
MDI observations of flares with different sized pupils, indicate that the 
energies quoted in Table~\ref{tab:table2} account for 80\,--\,95 per cent of the total, 
depending on the source distribution.}
compared with energy emitted in X-rays in the first 20 minutes of the
flare. It should be noted that the 3~mHz energy for the flares
preceding the 2005 January 15 flare are actually calculated at 3.5~mHz.
Extrapolating through the missing 4\,--\,5~mHz acoustic spectrum for
the flare of January 15, we project a total acoustic emission of
$\sim 4 \times 10^{20}$~J ($\sim 4 \times 10^{27}$~erg).

\subsection{Visible Continuum Emission}

Various aspects of visible continuum emission during the 2005 January
15 flare are shown in Figs.~\ref{fig:egpwrsnaps},
\ref{fig:acpwrsnaps} and \ref{fig:plots}.  The visible-continuum
images in Figs.~\ref{fig:egpwrsnaps} and \ref{fig:acpwrsnaps} were
obtained by MDI at 00:00~UT, $\sim$37~min before the onset of the
flare. We obtained visible continuum maps of AR10720 during the flare
from the GONG observatory at Mauna Loa. Technically, the GONG
``continuum intensity maps'' represent a measure of radiation in a
$\sim$1~\AA\ bandpass centred on the Ni~I~6768~\AA\ line, 
whose equivalent width is only a fraction of an \AA\ .  Frames c) and
d) in Figs.~\ref{fig:egpwrsnaps} and \ref{fig:acpwrsnaps} show the
difference in continuum intensity between the GONG images 30 seconds
before and after at the time indicated above the frame.  Continuum
emission is elongated along the magnetic neutral line, corresponding
closely to the lenticular component of seismic emission seen at
00:42~UT in Fig.~\ref{fig:egpwrsnaps}h.  The brightest emission seen
in the intensity difference shown in Fig.~\ref{fig:egpwrsnaps}d
comes from a very compact kernel whose location coincides very closely
with that of the conspicuous kernel of 6~mHz emission
(Fig.~\ref{fig:egpwrsnaps}h).

If we assume that the continuum emission emanates isotropically from an opaque surface%
\footnote{The assumption is that the specific intensity is independent
of the vantage, which implies that the total intensity decreases in
proportion to the cosine of the angle of the vantage from normal as a
result of foreshortening. If the source was assumed to be optically
thin, the resulting energy estimate would be greater by a factor of
two. The former appears to be the more realistic estimate for the
fraction of visible continuum radiation coming directly from the
chromosphere, based on the thesis that ionization of chromospheric
hydrogen at the temperature minimum renders the low chromosphere
opaque.}  the resulting estimate of the total energy emitted in the
visible continuum is $2.0 \times 10^{23}$~J ($2.0 \times
10^{30}$~erg).  This is $\sim$500 times the total seismic energy we
estimate the flare to have emitted into the holographic pupil.
Continuum radiation into the neighbourhood of the 6~mHz kernel
signature was $6.0 \times 10^{22}$~J ($6.0 \times 10^{29}$~erg).  This
accounted for $\sim$30 per cent of the total, as compared to 45 per cent of the
6~mHz seismic signature.  Continuum emission from in the neighbourhood
of the 6~mHz kernel was significantly more sudden than that of the
remainder of the acoustic signature.

The 2005 January 15 flare contributes to recent findings that relatively
small flares can emit disproportionate amounts of acoustic energy 
\citep{donea2005}. However, even in these cases the fraction of the energy that is released by the flare into the solar interior acoustic spectrum remains relatively small.

\subsection{The Surface Ripples}

Holography allows us to image the acoustic source of the sunquake when the 
surface manifestation of the seismic emission is difficult to detect. 
In the case of the exceptionally powerful seismic transient from the flare 
of 2005 January 15, the surface signature is quite evident in the raw 
MDI Doppler observations, a point to which A. Kosovichev (2005, private 
communication) drew our attention shortly after we reported the discovery 
of the sunquake to him.
To extract the seismic oscillations in the observations we subtracted 
consecutive MDI Doppler images separated by one minute in time.
We applied this Doppler-difference method to a period of observation 
($\sim$1 hour) around the time of the flare. 
Results are shown in Figs.~\ref{fig:ripples1} and \ref{fig:ripples2}. 

The Doppler signature of the flare is clearly evident at 00:40~UT
(Fig.~\ref{fig:ripples2}, left panel).  At approximately 20 minutes
after the appearance of the flare signature in the sunspot photosphere
(at 01:00~UT), we are able to see the seismic response of the
photosphere to the energy deposited by the flare in the form of
``ripples'' on the solar surface.  In the sequence of one-minute
Doppler-difference images in Fig.~\ref{fig:ripples1}, we can see the
asymmetrical ring-shaped wave packet propagating from the flare site
with the first wave-crest appearing approximately 12\,--\,15~Mm from
the flare in a North-Easterly direction.  The lower half of the
wave-packet has a much smaller amplitude and is propagating in a
South-Westerly direction. The arrows in Fig.~\ref{fig:ripples1}
indicate the location of the observed wave fronts. The
Doppler-difference images in Fig.~\ref{fig:ripples2} show a close-up
of the active region at the time of the flare (at 00:40:30~UT, left
panel) and the resulting ring-shaped wave packet (at 01:05:00 UT,
right panel).

The wave-packet was seen to propagate to a maximum distance of
approximately 21~Mm from the flare signal, hence travelling a total
distance of 6\,--\,9~Mm and lasting for about 8 minutes on the
surface, after which the wave amplitude dropped rapidly and the
disturbance became submerged in the ambient noise.  The lower half of
the wave-packet (propagating towards the South\,--\,Western part of
the active region, indicated by the lower of the two arrows
superimposed on the Doppler-difference images in Figs.~\ref{fig:ripples1} and \ref{fig:ripples2}) was much smaller in
amplitude and obscured for much of the 8 minutes.
\begin{figure}
\begin{center}
\includegraphics[width=0.49\textwidth, angle=0]{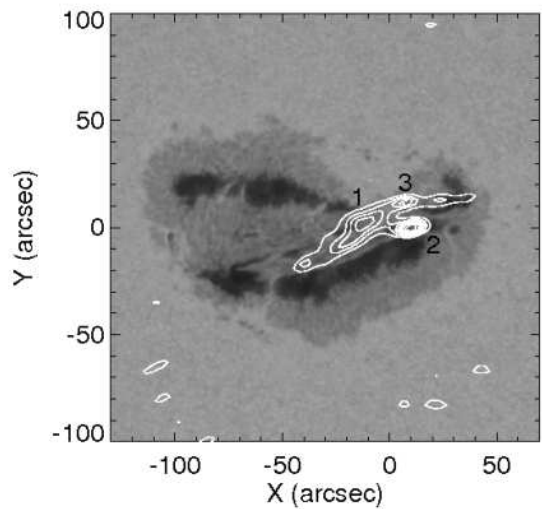}
\caption{\textit{TRACE} white-light image for AR10720, on 2005 January 15 (00:17:54~UT) with 
the 12-25 keV \textit{RHESSI} contours (10, 20, 30, 40, 50, 60 and 80 per cent of 
the maximum flux). 
The (0,0) coordinates correspond to the location of the seismic source.}%
\label{fig:trace}
\end{center}
\end{figure}

\begin{figure*}
\begin{center}
\includegraphics[width=0.7\textwidth, angle=0]{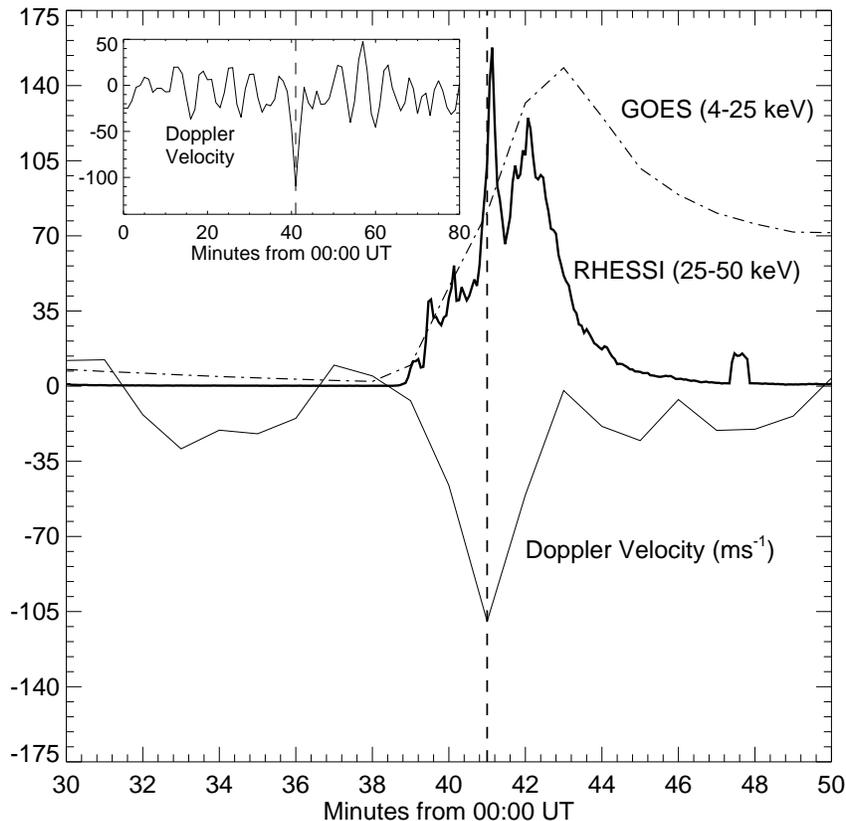}
\caption{The HXR flux in the 4\,--\,25 (0.5\,--\,4~\AA) and 25\,--\,50~keV energy ranges as observed by \textit{GOES} 
(dot-dashed curve, multiplied by a factor of $10^{6.5}$~Wm$^{-2}$) and \textit{RHESSI} (bold curve, 
multiplied by a factor of $10^{-2.8}$~counts) respectively. The solid curve represents 
the 1-minute mean averages of the Doppler velocity (ms$^{-1}$) in the quake region for 
the period 00:00:00\,--\,01:20:00~UT. 
The dashed vertical line represents the observed maximum emission at 6~mHz.}
\label{fig:Xrays}
\end{center}
\end{figure*}
 
\subsection {Hard X-Ray Emission}

The \textit{TRACE} data for the 2005 January 15 flare in the white-light channel
have a variable cadence for the period 00:00:00\,--\,01:00:00~UT.
Fig.~\ref{fig:trace} shows the \textit{TRACE} white-light image taken at
00:17:54~UT, approximately 10~minutes before the onset of the X1.2
flare with the \textit{RHESSI} 12-25~keV contours overlaid. The \textit{RHESSI} HXR
image is averaged over the period 00:41:33\,--\,00:42:34~UT.  The time
of peak intensity in this energy band occurs at 00:42:04~UT, a close
temporal correlation with the maximum of the seismic emission detected
at 6~mHz.  The HXR emission is thought to represent bremsstrahlung
emission from high-energy coronal electrons impinging into the
chromosphere \citep{b1971}.

The 12\,--\,25~keV emission at 00:42:00~UT extends along the neutral
magnetic line. We identify three compact HXR sources (see the numbers
in Fig.~\ref{fig:trace}).  Source 2 is the strongest, while source 3
is the weakest.  These could represent the foot-points of a complex
magnetic loop.  However, source 1 (which emits 50 per cent of the total flux)
spatially coincides with the lenticular component of the 6~mHz seismic
source (see Fig.~\ref{fig:egpwrsnaps}).  This reinforces the role of
non-thermal particles in supplying the energy that drives the seismic
emission.  Similar comparisons have been observed in other flares \citep{dl2005}.

Furthermore, Fig.~\ref{fig:Xrays} reveals that the velocity impulse
of the flare in the sunspot photosphere was almost as sharp as the HXR
flux detected in the 4\,--\,25~keV (0.5\,--\,4~\AA) energy range by
the \textit{GOES} satellite, but the maximum HXR emission (observed at
$\sim$00:43:00~UT) appears to have occurred $\sim$2 minutes after the
maximum velocity depression at the photosphere (00:41:00 UT). In
fact, a sudden drop of approximately 100~ms$^{-1}$ in the mean
velocity of the Doppler signal (an upflow) is observed in the 3 minute
period from 00:38:00-00:41:00~UT.  The \textit{RHESSI} HXR peak in the higher
energy band of 25\,--\,50~keV plotted in Fig.~\ref{fig:Xrays} occurs
at $\sim$00:41:00~UT, which temporally coincides with both the maximum
of the seismic source at 6~mHz and the velocity depression at the
photosphere. We also note that the peak emission in the 3\,--\,12~keV
energy band detected by both \textit{GOES} (1\,--\,8~\AA) and \textit{RHESSI} (occurring
at $\sim$00:44:00 and $\sim$00:47:00~UT respectively, but not
plotted), also have a close temporal correlation with the maximum of
the seismic emission.

\section{Discussion and Summary} 

The X1.2-class flare of 2005 January 15 produced one of the most
powerful sunquakes detected to date and by far the most conspicuous,
on account of exceptionally powerful emission above 5~mHz from a
compact source.  Certain qualities exhibited by the flare of January
15 are shared by all other known acoustically active flares.  The
first is the coincidence between strong compact acoustic sources and
nearby signatures of HXR emission.  This suggests that high-energy
particles supply the energy that drives the acoustic emission, and it
is evident from the electromagnetic emission attributed to these
particles that they contain more than sufficient energy for this
purpose.  The appearance of sudden, conspicuous white-light emission
from the flare of 2005 January 15 closely co-spatial with the location and
morphology of the holographic signatures is similarly characteristic
of all other known acoustically active flares so far.

\citet{kz1998} proposed that seismic emission into the solar interior
in sunquakes is the continuation of a chromospheric shock and
condensation resulting from explosive ablation of the chromosphere and
propagating downward through the photosphere into the underlying solar
interior. Chromospheric shocks are well known under such
circumstances, based on red-shifted H$\alpha$ emission at the flare
site at the onset of the flare.  The theory of their dynamics was
worked out at length by \citet{fisher1985a,fisher1985b,fisher1985c} and
others since.  The hypothesis that solar interior emission is a direct
continuation of such shocks was considered by \citet{dl2005}, who found
the signature of a strong downward-propagating chromospheric transient
in Na~D$_1$-line observations of the flare of 2003 October 29.
However, we are now aware of similar chromospheric transient
signatures with no significant attendant holographic signature to
indicate seismic emission into the active region
subphotosphere\footnote{An example is seen in the western foot-point of
the magnetic loop that hosted 2003 October 29 flare (see right frame
in second row of fig.~8 labelled ``Red \hbox{[0 min]}'' in
\citet{dl2005}).  The corresponding signature of sudden white-light
emission, seen in fig.~9 of the same, shows only a weak signature at
the same location.  The seismic signature, seen in the lower left
corner of the upper right frame of fig.~11 of the same, shows
correspondingly weak seismic emission.}%

In these instances, the signature of sudden white-light emission is
relatively weak.  Following \citet{machado1989}, \citet{donea2005}
proposed to attribute the lack of seismic emission where there is a
strong chromospheric transient but only a weak or absent white-light
signature to strong radiative damping that depletes the chromospheric
transient before its arrival into the low photosphere.

In all acoustically active flares encountered to date, there is a
strong spatial correlation between the sources of seismic emission and
sudden white-light emission.  This remains conspicuously the case for
the flare of 2005 January 15, as a comparison between Figs.~1d and 1h
shows.  In some instances, e.g. the large flares of 2003 October 29,
the source of the white-light emission has been much more extensive
than the source of the acoustic emission, the former many times the
area of the latter and encompassing it.  However, in these instances
the temporal profile of visible continuum emission significantly away
from any of the sites of seismic emission has been comparatively
sluggish and diffuse.  What has particularly and consistently
distinguished the white-light signature in the neighbourhood of the
acoustic emission has been the suddenness of its appearance, on a time
scale of a minute or two, and possibly considerably less than a minute
given that the observations of continuum emission associated with
flares to date have been limited to a cadence of one minute.

It should be kept in mind that the energies released in known seismic
transients have invariably been a small fraction of the energy
released into the visible continuum spectrum. The actual fraction has
varied considerably, from a few millionths, in the case of the flare
of 2003 October 29 \citep{dl2005}, to a few thousandths, for the flare
of 2005 January 15.  However, if only the sudden-onset continuum
emission in the neighbourhood of the seismic sources is included, then
the ratio for the flare of 2003 October 29 is similar to that of the
flare of 2005 January 15.  This is what is listed in
Table~\ref{tab:table2} of this study.

{\it The close coincidence between the locations of sudden white-light
emission and seismic transient emission in all acoustically active
flares to date suggests that a substantial component of the seismic
emission seen is a result of sudden heating of the low photosphere
associated with the visible continuum emission seen.}  A complete
analysis of wave emission as a result of transient heating involves
detailed considerations of energy and momentum balance.  An
approximate account of these was undertaken by \citet{donea2005}.
Basic considerations of momentum balance are described in Section 4.3
of \citet{donea2005}, adapting the discussion by
\citet{canfield1990} of momentum balance in chromospheric
transients to transients similarly excited by sudden heating in the
low photosphere.

\citet{donea2005} devised a rough, preliminary physical model to
estimate the energy of the seismic transient to be emitted as a result
of sudden, momentary heating of the low photosphere to a degree
consistent with the transient white-light signature closely coincident
with the seismic source in the M9.5-class flare of 2001 September 9.
Their estimate expressed the energy, $E$, of the seismic transient in
terms of the thermal energy, $U$, radiated or dissipated into the low
photosphere, and the fractional increment, $\delta p/p$, in pressure
that would result from the heating:
\begin{equation}
E ~=~ {1 \over 2} H {(\delta p)^2 \over p}
  ~=~ {1 \over 3} ~{\delta p \over p} ~\delta U,
\label{eq:EdUdp}
\end{equation}
where $H$ is the $e$-folding height of the photospheric density.
This relation appears to be roughly consistent with the few-percent continuum
intensity variations observed for the flare of 2001 September 9, if the relation 
between $\delta p$ and $\delta I$, the variation in continuum intensity, can
be approximated by the Stefan-Boltzmann law,
\begin{equation}
{\delta p \over p} ~=~ {\delta T \over T}
             ~=~ {1 \over 4}{\delta I \over I},
\label{eq:boltzmann}
\end{equation}
and the heating is accomplished within a duration not excessively
longer than $\tau _{ac} = 1/\omega _{ac} \sim 40$~s, where
$\omega_{ac}$ is the acoustic cutoff frequency in the low photosphere
(see Section 4 of \citet{dl2005}).  A similar exercise applied to
the flare of 2005 January 15 leads to similar results.  In fact, the
ratio of the seismic energy to the electromagnetic energy is roughly
the same for both of these flares, as are the mean intensity
increments if credible boundaries are chosen over which to take the
mean.  To the extent that we can resolve the fine details, acoustic
emission from the flare of 2005 January 15 could reasonably be the
result of photospheric heating similar to that of the 2001 September 9
flare but over approximately twice the area.  Differences between the
two flares could be attributed to differing photospheric or
subphotospheric thermal conditions and differing magnetic fields, for
which the foregoing approximation contains no account.

A detailed examination of the physics of heated magnetic photospheres
is needed to lend credibility to the hypothesis that seismic emission
from acoustically active flares is driven by sudden heating of the low
photosphere by any mechanism whatever.  At this point we will only say
that this hypothesis appears to be consistent with our present limited
understanding of the observations.  However, there is some controversy
as to the implications of visible continuum emission during flares
with respect to heating of the low photosphere.  In the case of the
flares of 2003 October 28\,--\,29, the signature of high-energy protons
along with the particles that gave rise to X-ray emission lent
considerable weight to the interpretation of visible continuum
emission in terms of a heated low photosphere, as protons are
sufficiently massive to penetrate to the bottom of the photosphere and
heat it directly by collisions.  The flare of 2005 January 15, on the
other hand, confronts us with an instance of intense seismic emission
with no indication of high-energy protons among the energetic
particles that supply the energy on which the acoustic emission
depends.  Energetic electrons consistent with HXR signatures cannot
penetrate into the low photosphere in anywhere near sufficient numbers
to account for the heating required by the seismic signatures
\citep{metcalf1990}. \citet{chen05} also affirm that the white-light flare
signatures highlight the importance of radiative back-warming in transporting
the energy to the low photosphere when direct heating by beam electrons is impossible.

In such cases, it appears to be well established that the origin of
white-light emission would have to be entirely in the chromosphere,
where energetic electrons dissipate their energy
\citep{metcalf1990,zharkova1991,zharkova1993}, mainly by ionizing
previously neutral chromospheric hydrogen approximately to the depth
of the temperature minimum.  Nevertheless, even in these instances, it
appears that the low photosphere itself would be significantly heated
as a secondary, but more or less immediate, effect of chromospheric
ionization.  This is primarily the result of Balmer and Paschen
continuum edge recombination radiation from the overlying ionized
chromospheric medium, approximately half of which we assume radiates
downward and into the underlying photosphere.  When the intensity,
$\delta I$, of this downward flux is commensurate with a temperature
perturbation, $\delta T$, consistent with the Stefan-Boltzmann law
(equation [\ref{eq:boltzmann}]), the result of such a flux is
understood to be heating of the low photosphere such as to bring about
a temperature increment of roughly this order within a few seconds
\citep{donea2005,machado1989, metcalf2003}.  Heating of the photosphere
by the mechanism described above is known as ``back-warming''
\citep{metcalf2003} and a substantial fraction of the continuum
emission seen in white-light flares is thought to represent the
downward flux from an ionized chromosphere thermally re-emitted by the
now heated photosphere.  In this light, the strong correlation between
sources of white-light and seismic emission into the solar interior
might be regarded as strong support for the back-warming hypothesis
when this relation persists in flares devoid of protons among the
high-energy particles that drive the flare.  This is certainly the
case for the flare of 2005 January 15.

\citet{dl2005} and \citet{donea2005} summarize our understanding of the relationship 
between the efficiency of seismic emission and the suddenness of the heating 
that drives the seismic transient.
Based on these considerations, one has to suspect that the perceptibly more sudden 
profile of continuum emission in the neighbourhood of the kernel component of the 
6~mHz emission accounts to a significant degree for the disproportionate power in 
the 6~mHz egression-power signature.
This is one of the many aspects of flare acoustics that would benefit from detailed 
modelling, including a careful account of magnetic forces.

At this point, our understanding of seismic emission from flares remains
relatively superficial.
However, evidence for the general involvement of photospheric heating is
now considerable.
What is needed for further understanding is detailed modelling with a careful
account of the physics, including radiative transfer and magnetic forces
in realistic sunspot photospheres and subphotospheres.
With such an understanding, acoustic emission from flares could contribute major 
benefits to seismic diagnostics of active region subphotospheres and the physics 
of mode conversion \citep{cally2000}.

%

\section*{Acknowledgments}

We have benefited greatly from the insights of Drs. T. Metcalf and V. Zharkova.
We also greatly appreciate comments by Dr. A. McClymont.
Dr. A. Kosovichev drew our attention to certain interesting aspects of the 
surface ripples caused by the seismic transient, of which the holographic
signatures represent the sources rendered in Figs.~1g,h.
The research reported here heavily utilized data obtained by the MDI instrument 
on the \textit{SOHO} spacecraft, operated by the National Aeronautics and Space 
Administration (NASA) and the European Space Agency (ESA).
This work further utilized data obtained by the GONG++ project, managed by the 
National Solar Observatory, a Division of the National Optical Astronomy 
Observatories, which is operated by AURA, Inc. under a cooperative agreement 
with the National Science Foundation.  
This research was supported by grants and contracts from the Astronomy and 
Stellar Astrophysics Branch of the National Science Foundation.
It was also supported by NASA's Sun-Earth Connection/Solar Heliospheric Physics 
Program.


\begin{thebibliography}{}

\bibitem[\protect\citeauthoryear{Braun}{1995}]{braun1995} 
Braun D. C., 1995, ApJ, 451, 859

\bibitem[\protect\citeauthoryear{Braun, Duvall \& LaBonte}{1988}]{bdl1988}
Braun D. C., Duvall T. L. Jr., LaBonte B. J., 1988, ApJ, 335, 1015

\bibitem[\protect\citeauthoryear{Braun et al.}{1998}]{blff1998}
Braun D. C., Lindsey C., Fan Y., Fagan M., 1998, ApJ, 502, 968

\bibitem[\protect\citeauthoryear{Braun \& Lindsey}{1999a}]{bl1999a}
Braun D. C., Lindsey C., 1999a, ApJ, 510, 494

\bibitem[\protect\citeauthoryear{Braun \& Lindsey}{1999b}]{bl1999b}
Braun D. C., Lindsey C., 1999b, ApJ, 513, L79

\bibitem[\protect\citeauthoryear{Brown}{1971}]{b1971}
Brown J. C., 1971, Solar Phys., 18, 489

\bibitem[\protect\citeauthoryear{Cally}{2000}]{cally2000} 
Cally P. S., 2000, Solar Phys., 192, 395

\bibitem[\protect\citeauthoryear{Canfield et al.}{1990}]{canfield1990}
Canfield R. C., Zarro D. M., Metcalf T. R., Lemen J. R., 1990, Solar Phys., 348, 333

\bibitem[\protect\citeauthoryear{Chen \& Ding}{2006}]{chen05} 
Chen Q. R., Ding M. D., 2006, ApJ , 641, 1217

\bibitem[\protect\citeauthoryear{DeForest}{2004}]{deforest2004}
DeForest C. E., 2004, Solar Phys., 219, 3

\bibitem[\protect\citeauthoryear{Donea, Braun \& Lindsey}{1999}]{dbl1999} 
Donea A. -C., Braun D. C., Lindsey C., 1999, ApJ, 513, L143  

\bibitem[\protect\citeauthoryear{Donea \& Lindsey}{2005}]{dl2005} 
Donea A. -C., Lindsey C., 2005, ApJ, 630, 1168

\bibitem[\protect\citeauthoryear{Donea et al.}{2005}]{betal2006} 
Donea A. -C., Besliu D., Cally P. S., Lindsey C., 2005, in Leibacher J., 
Uitenbroek H., Stein B., eds, ASP Conf. Ser., Solar MHD: Theory and Observations - A 
High Spatial Resolution Perspective, Astron. Soc. Pac., in press

\bibitem[\protect\citeauthoryear{Donea et al.}{2006}]{donea2005}
Donea A. -C., Besliu-Ionescu D., Cally P. S., Lindsey C., Zharkova V. V., 2006, Solar Phys., accepted June 2006


\bibitem[\protect\citeauthoryear{Fisher, Canfield \& McClymont}{1985a}]{fisher1985a}
Fisher G. H., Canfield R. C., McClymont A. N., 1985a, ApJ, 289, 414   

\bibitem[\protect\citeauthoryear{Fisher, Canfield \& McClymont}{1985b}]{fisher1985b} 
Fisher G. H., Canfield R. C., McClymont A. N., 1985b, ApJ, 289, 425  
 
\bibitem[\protect\citeauthoryear{Fisher, Canfield \& McClymont}{1985c}]{fisher1985c} 
Fisher G. H., Canfield R. C., McClymont A. N., 1985c, ApJ, 289, 434   

\bibitem[\protect\citeauthoryear{Hudson}{1972}]{hudson1972} 
Hudson H. S., 1972, Solar Phys., 24, 414

\bibitem[\protect\citeauthoryear{Kosovichev \& Zharkova}{1998}]{kz1998} 
Kosovichev A. G., Zharkova V. V., 1998, Nature, 393, 317

\bibitem[\protect\citeauthoryear{Lindsey and Braun}{1999}]{lb1999} 
Lindsey C., Braun D. C., 1999, ApJ, 510, 494

\bibitem[\protect\citeauthoryear{Lindsey and Braun}{2000}]{lb2000}
Lindsey C., Braun D. C., 2000, Solar Phys., 192, 261

\bibitem[\protect\citeauthoryear{Machado et al.}{1989}]{machado1989}
Machado M. E., Emslie A. G., Avrett E. H., 1989, Solar Phys., 124, 303

\bibitem[\protect\citeauthoryear{Metcalf, Canfield \& Saba}{1990}]{metcalf1990} 
Metcalf T. R., Canfield R. C., Saba J., 1990, ApJ, 365, 391

\bibitem[\protect\citeauthoryear{Metcalf et al.}{2003}]{metcalf2003}
Metcalf T. R., Alexander D., Hudson H. S., Longcope D., 2003, ApJ, 595, 483

\bibitem[\protect\citeauthoryear{Scherrer et al.}{1995}]{setal1995} 
Scherrer P. H. et al., 1995, Solar Phys., 162, 129

\bibitem[\protect\citeauthoryear{Zharkova \& Kobylinskii}{1991}]{zharkova1991}
Zharkova V. V., Kobylinskii V. A., 1991, Sov. Astron. Lett., 17,  34

\bibitem[\protect\citeauthoryear{Zharkova \& Kobylinskii}{1993}]{zharkova1993} 
Zharkova V. V., Kobylinskii V. A., 1993, Solar Phys., 143, 249

\end{thebibliography}
\end{document}